\documentclass[11pt,a4paper]{article}
\usepackage[hyperref]{emnlp2018}
\usepackage{times}
\usepackage{latexsym}
\usepackage{graphicx}
\usepackage{kotex}
\usepackage{url}
\usepackage{todonotes}
\usepackage{tabularx}
\usepackage{booktabs}

\usepackage{color}
\usepackage{soul}

\aclfinalcopy 

\title{Football and Beer - a Social Media Analysis on Twitter in Context of the FIFA Football World Cup 2018}

\author{Roland Roller, Philippe Thomas, Sven Schmeier \\
	    Language Technology Lab, DFKI,\\
		Berlin, Germany\\
	    {\tt \{firstname.surname\}@dfki.de}}

\date{}

\begin{document}
\maketitle
\begin{abstract}
In many societies alcohol is a legal and common recreational substance and socially accepted. Alcohol consumption often comes along with social events as it helps people to increase their sociability and to overcome their inhibitions. On the other hand we know that increased alcohol consumption can lead to serious health issues, such as cancer, cardiovascular diseases and diseases of the digestive system, to mention a few. This work examines alcohol consumption during the FIFA Football World Cup 2018, particularly the usage of alcohol related information on Twitter. For this we analyse the tweeting behaviour and show that the tournament strongly increases the interest in beer. Furthermore we show that countries who had to leave the tournament at early stage might have done something good to their fans as the interest in beer decreased again.

\end{abstract}

\section{Introduction}

Alcohol can lead to serious health issues. For instance, even though there is no apparent threshold, even one drink of alcohol per day on average can significantly increase the risk of cancer \cite{Roerecke2012}.

Studies have shown, that the exposure to media and commercial communications on alcohol is associated with the likelihood that adolescents will start to drink alcohol, and with increased drinking amongst baseline drinkers \cite{Anderson:2009}. In addition to that social events can have a influence on drinking behaviour. In course of this \newcite{Curtis:2018} apply a Twitter analysis and show that topics such as sporting events, art and food-related festivals are positively correlated to alcohol consumption on US county level. 

Various other studies have also explored alcohol-related content on social media, particularly Twitter. \newcite{Abbar:2015} carry out a food analysis on Twitter and identify weekly periodicities in context of daily volume of tweets mentioning food. Moreover, authors show a correlation between state obesity and caloric value of food (also alcoholic beverages). Instead \newcite{Culotta:2013} analyse alcohol sales volume in context of Twitter messages. \newcite{Kershaw:2014} investigate regional alcohol consumption patterns in the UK, while \newcite{Hossain:2016} explore alcohol consumption patterns in various areas in the US. \newcite{Curtis:2018} target the prediction of excessive drinking rates and \newcite{Huang:2017} examine alcohol- and tobacco-related behaviourial patterns. And finally, \newcite{Moreno:2010} carry out a content analysis of adolescents on social media. 

This work examines alcohol consumption on Twitter in context of the FIFA Football World Cup 2018. We make use of the results of the above mentioned works, especially the observed correlation between people's behaviour on Twitter and in their real life in context of consumption. The study is carried out across all participating countries of the tournament and explores the influence of the event on the drinking behaviour of people. 









\section{Experimental Setup}\label{setup}

The FIFA World Cup 2018 was taking place from 14th of June until 15th of July. Within the group stage 32 participating teams were playing in 8 groups and completed 3 matches each. After that the two best teams of each group went to the knockout stage. 
A match usually lasts 90 minutes plus 15 minutes of break ($\approx$2 hours). This work analyses the tweeting behaviours during a match. In the following a match is defined as a time period of one hour before kick-off and three hours after the kick-off. The hour before and after the game are included as supporters might express excitement for the game.


\subsection{Data Collection}

Tweets over a period between 05/31/2018 and 07/23/2018 were collected, which covers the period of the tournament, but also two weeks before and one week after. As this work analyses messages from all participating countries of the tournament, messages were crawled containing various emojis due to their language independence. The considered emojis are listed in Table \ref{table:emojis}. In the rest of the work we refer to them as \textit{BEER}, \textit{WINE}, \textit{SAKE} and \textit{BALL}.

\begin{table}[!th]
  \centering
  \begin{tabular}{ rrrr }
\includegraphics[width=0.045\textwidth]{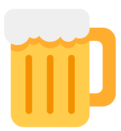} & \includegraphics[width=0.045\textwidth]{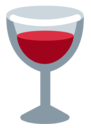} & \includegraphics[width=0.045\textwidth]{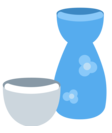} & \includegraphics[width=0.045\textwidth]{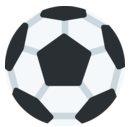}
\\
\end{tabular}
\caption{Emojis used for Twitter crawling}
\label{table:emojis}
\end{table}

\vspace{-0.4cm}


\subsection{Country Assignment}

In this work, Tweets are examined according to the different participating countries. As only a small number of Tweets contain country related information (8.63\%), Tweets lacking this information had to be assigned automatically to the corresponding country of each user. A classifier was trained based on the approach of \newcite{Thomas:2017}, which is able to detect the origin of a Tweet based on text and meta information. 


As sanity check, Tweets which actually contain information about its origin were compared to the automatically assigned country. On those messages the model achieves an accuracy of above \textbf{91}\%. 

\subsection{Preprocessing}

Collected Tweets were then mapped to small time intervals of one hour, according to the target label (e.g. beer emoji). For instance a message sent on 06/27/2018 at 5:25 pm (UTC) and containing a beer emoji is assigned to its country, then assigned to the set of beer emoji Tweets, and finally mapped to the time interval 06/27/2018, 5:00 pm. 
All messages of a particular label and a particular time interval are summed up. In this way a list for each target label is generated containing \textit{time intervals}, \textit{country origin} and \textit{number of relevant Tweets} for this interval. For the following analysis these lists are used as input. One line is considered as TCF (time-country-frequency) triple.

\section{Analysis}\label{analysis}

In the following the different Tweets containing the target emojis are analysed in detail. All examinations which include significance testing use one-sided paired t-test.

\subsection{Outside the Tournament}

\begin{figure}[bht!]
\centering
\includegraphics[width=0.43\textwidth]{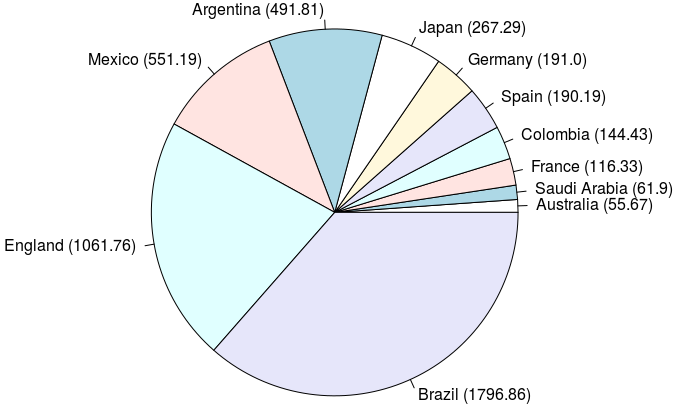}
\caption{Average number of alcohol related (BEER) Tweets per day outside the tournament}
\label{figure:pie_avg}
\end{figure}

Figure \ref{figure:pie_avg} shows the average number of Tweets per day containing \textit{BEER} emojis before/after the tournament from the participating countries. Participants with less than 50 \textit{BEER} Tweets per day are excluded, to show more meaningful results. We refer to this group as \textit{avg50}. In average, more than half of the \textit{BEER} Tweets per day come from Brazil and England together. Results show no direct correlation to the statistics \textit{Harmful use of alcohol}\footnote{\url{http://apps.who.int/gho/data/node.sdg.3-5-viz}, accessed 19.07.2018} of the WHO. The reasons for this can only be possibly found out by a longterm analysis going further than pure statistics. In that list Brazil with 7.8 litres of pure alcohol per capita is actually ranked further to the end. Instead countries such as Germany (13.4), France (12.6), England (11.4) and Australia (10.6) would be ranked to the top. 

In order to draw a fair comparison to \textit{pure alcohol per capita} the number of active Twitter users must be taken into account. It turns out that dividing the avg. \textit{BEER} Tweets by the number of active Twitter users does not change much. Argentina switches place with England, Saudi Arabia moves to the very end and Japan drops just in front of it. Columbia moves slightly up. 

\begin{figure}[bht!]
\centering
\includegraphics[width=0.49\textwidth]{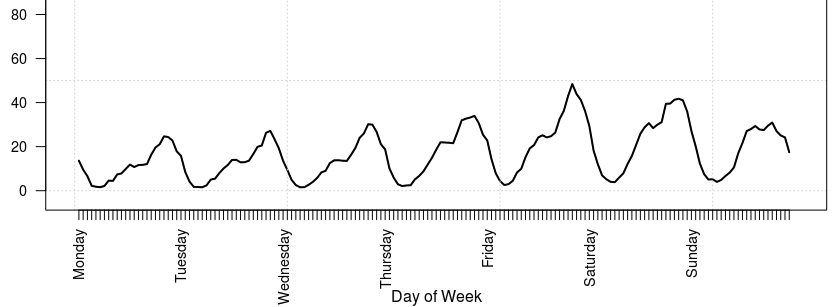}
\caption{Mean of all alcohol related (\textit{BEER}) Tweets before/after the tournament from \textit{avg50}}
\label{figure:tweet_vis}
\vspace{-0.5cm}
\end{figure}

Next mean is calculated for all \textit{BEER} Tweets from \textit{avg50} for each day of the week. The resulting graph is presented in Figure \ref{figure:tweet_vis} and visualises, similarly to \newcite{Abbar:2015}, particular periodicities. Firstly single days can be recognised as small peaks. Moreover, towards the end of the week, peaks are slightly increased compared to the beginning of the week. Using this data it is for instance possible to deduct, that \textbf{people tweet significantly more about alcohol on the evening} (from 4pm until 1am) (p$<$0.001). 
Moreover, the data also shows, that \textbf{people tweet significantly more about alcohol at the weekend} (Friday 4pm - Monday 6am) (p$<$0.001). %
in comparison to the rest of the week. 


\subsection{The Tournament}

In this subsection alcohol related Tweets during the tournament are examined. The first question to address is whether supporters of their national team tweet more during the match in comparison to other periods. Reference periods are the days after each match during the same time slots.  

The analysis shows that \textbf{people from 19 countries use \textit{BEER} significantly more when their team is playing} (p$<$0.05, 10 of them with p$<$0.001). 
Among the 13 other countries, only Japan and Saudi Arabia are from \textit{avg50}. Interestingly Croatia, which reached the final, does not show any significant increase, but the general usage of \textit{BEER} is generally very low here. Considering \textit{WINE} Tweets, only Brazil, Poland and Belgium and for \textit{SAKE} only Mexico show a significant increase in Tweets during the matches of their team (p$<$0.05). However the number of Tweets are low in comparison to \textit{BEER}. 

\begin{figure}[bht!]
\centering
\includegraphics[width=0.45\textwidth]{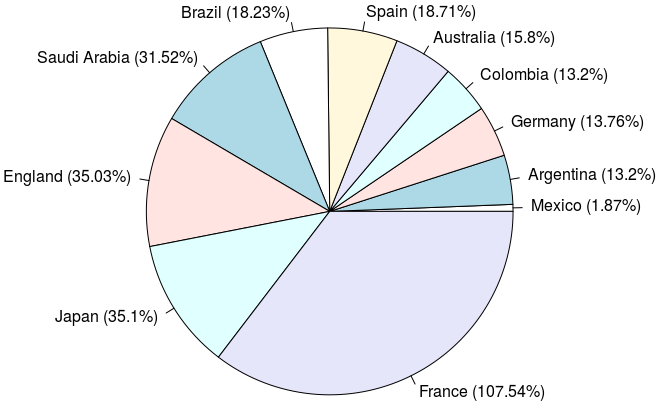}
\caption{Increase of beer related Tweets per day during World Cup, until leaving the tournament}
\label{figure:pie_increase}
\end{figure}

Figure \ref{figure:pie_increase} presents an overview on how the tournament influences the average usage of \textit{BEER} per day of \textit{avg50}, while the team is in the tournament. France shows the largest increase of BEER Tweets per day in average of more than 107\%, followed by Japan with 35\%. The increase from Japan is surprising as Japanese people do not tweet significantly more during the matches of their team. Possible explanations might be that matches are broadcasted often late in the evening due to the time difference to Russia. For this reason people might meet up earlier, thus start drinking earlier. Another explanation can be just the fact that there is a high interest for the tournament in general in Japan. 

Brazil instead shows `only' an increase of 18\%. Even with this small increase (in comparison to others) Brazil remains the country with the largest number of avg. \textit{BEER} Tweets per day. Considering all countries Morocco has the strongest increase (407.5\%) and also the strongest decrease after leaving the tournament (-68.97\%). On the other hand, outside the tournament Morocco has a very low number of \textit{BEER} Tweets per day (0.48), so the increase might be not too serious. Interestingly only Peru shows a decrease
 during the tournament (-16.4\%). Considering the \textit{avg50}, Colombia and Brazil showed the strongest decrease when their team left the tournament with -18.81\% and -10.59\% respectively.  

Generally the results show, that from almost all countries more alcohol related Tweets can be found during the tournament. Moreover, in most cases the avg. number of \textit{BEER} Tweets  decreases when the team leaves the tournament. However, in some cases an increase in Tweets can be detected. Senegal for instance increases the number of \textit{BEER} Tweets up to 54\%, followed by Uruguay (28\%) and Australia (10\%).

\subsection{Top-5 Matches}

This subsection analyses the different matches of the tournament for popularity in terms of \textit{BEER} and \textit{BALL}. In order to have a fair comparison data is normalized first. The average number of Tweets of each country outside the tournament is subtracted from the number of Tweets during the tournament at a given time and day. 

Table \ref{table:top_matches} presents the Top-5 matches involving \textit{BEER} and \textit{BALL}. The table shows that more people use the football emoji than the beer emoji. In terms of \textit{BALL}, the Top-5 list contains the final (France-Croatia), the opening (Russia-Saudi Arabia) and some other games involving recent European and World Champions.

\begin{table}[!th]
  \centering
  \small
\begin{tabular}{ llll }
\toprule
\multicolumn{1}{c}{\includegraphics[width=0.03\textwidth]{twitter_pics/beer.png}} & \multicolumn{1}{c}{\#} & \multicolumn{1}{c}{\includegraphics[width=0.03\textwidth]{twitter_pics/football.png}} & \multicolumn{1}{c}{\#} \\
\midrule
Mexico-Sweden & 1313 & Portugal-Spain & 5100 \\
German-S. Korea &  &  &  \\\midrule
Brazil-Belgium  & 1305 & France-Croatia & 4637 \\\midrule
Serbia-Brazil & 1250 & Russia-S. Arabia  & 4253 \\
Switzerland-C. Rica & & &  \\\midrule
Nigeria-Iceland & 1104 & Germany-Mexico  & 3798 \\\midrule
Brazil-Costa Rica & 1092 & France-Argentina & 3616 \\
\bottomrule
\end{tabular}
\caption{Top-5 matches of the tournament in terms of beer and football emoji (normalized)}
\label{table:top_matches}
\end{table}

From \textit{BEER} perspective we find on the first and third palce matches which took place in parallel. Considering that, the most popular single match was Brazil-Belgium in the Quarter Final. Ranked 4th is Nigeria-Iceland, which is surprising, as both countries are not tweeting much about beer. Analysing the results in more detail reveals, that all 7 matches took place on a day Brazil played. Even though Tweets were normalized, the influence of Brazilian \textit{BEER} Tweets before and after a match of their team is enormous, so that even the Nigeria-Iceland match achieved a high rank.





%

%


%
\section{Results}\label{results}

This work presented a short analysis of alcohol related emojis in context of the FIFA football World Cup 2018. With the start of the tournament we showed, that most countries strongly increase the number of Tweets containing beer emojis. As many people tweet less after their national team left the tournament, we draw the conclusion that leaving the tournament early, as Germany did, is the healthiest solution -  unless you are Peru. We also showed that people of many participating teams of the tournament tweet significantly more about alcohol during a match of their team. Furthermore we presented the increase of alcohol related Tweets during the tournament and the most popular games in terms of beer and football emojis. Finally we showed, that Brazil tweets by far the most about beer. Cheers!

\section*{Acknowledgements}
This research was supported by the German Federal Ministry of Economics and Energy (BMWi) through the project MACSS (01MD16011F).
\vspace{-0.6cm}
\bibliography{emnlp2018}
\bibliographystyle{acl_natbib_nourl}

\end{document}